\UseRawInputEncoding
\documentclass[aps,prb,twocolumn]{revtex4-2}
\pdfoutput=1
\usepackage{dcolumn} 
\usepackage{grffile} 
\usepackage{bm}
\usepackage{subfigure}
\usepackage{amssymb}
\usepackage{amsmath}
\usepackage{graphicx}
\usepackage[pdftex]{hyperref}
\hypersetup{colorlinks=true,citecolor=blue,linkcolor=blue,urlcolor=blue}
\usepackage[all]{hypcap}
\usepackage{color}
\definecolor{red}{rgb}{0.8, 0.0, 0.0}
\definecolor{blue}{rgb}{0.06, 0.2, 0.65}
\definecolor{green}{rgb}{0.0, 0.6, 0.0}

\usepackage{xfp}

\begin{document}

\title{Registry dependent potential for interfaces of gold with graphitic systems}
\author{Wengen Ouyang$^1$, Oded Hod$^2$, Roberto Guerra$^3$}

\email{roberto.guerra@unimi.it, odedhod@tauex.tau.ac.il}

\affiliation{
  $^1$Department of Engineering Mechanics, School of Civil Engineering, Wuhan University, Wuhan, Hubei 430072, China
  \\$^2$Department of Physical Chemistry, School of Chemistry,
  The Raymond and Beverly Sackler Faculty of Exact Sciences and The Sackler Center for Computational Molecular and Materials Science, Tel Aviv University, Tel Aviv 6997801, Israel
  \\$^3$Center for Complexity and Biosystems, Department of Physics, University of Milan, 20133 Milan, Italy
}

\begin{abstract}
We present a semi-anisotropic interfacial potential (SAIP) designed to classically describe the interaction between gold and two-dimensional (2D) carbon allotropes such as graphene, fullerenes, or hydrocarbon molecules. The potential is able to accurately reproduce dispersion corrected density functional theory (DFT+D3) calculations performed over selected configurations: a flat graphene sheet, a benzene molecule, and a C$_{60}$ fullerene, physisorbed on the Au(111) surface. The effects of bending and of hydrogen passivation on the potential terms are discussed.
The presented SAIP potential provides a noticeable improvement in the state-of-the-art description of Au-C interfaces. Also, its functional form is suitable to describe the interfacial interaction between other 2D and bulk materials.
\end{abstract}

\maketitle

\section{Introduction}\label{sec.intro}

The reproducibility of phenomena emerging at the interface between two bodies in contact is often limited by the availability of clean and well controllable surfaces. Gold and graphene are two very stable materials that can be produced with a high level of crystallinity and cleanness over large surface areas.
For these reasons the gold/graphene interface has been the prototypical choice for a large number of case studies, including diffusing and sliding gold nanoclusters \cite{Luedtke1999, Lewis2000, Guerra2010, Dietzel2013, Cihan2016, Lodge2016}, nanomanipulation of graphene nanoribbons \cite{Kawai2016, Gigli2017, Gigli2018}, plasmon-enhanced optics \cite{Muszynski2008, Zhu2013}, sensing and biomedical applications  \cite{Turcheniuk2015}, surface-enhanced Raman scattering \cite{Goncalves2009, Song2019}, among many others.

Despite the vast interest expressed in this composite system by the scientific community, to date no reliable classical force field is available for computational simulations involving interfaces of gold with graphitic systems.
Since the interaction between graphene and metals is mainly governed by van der Waals forces \cite{Tesch2016, Forti2020}, a simple two-body Lennard-Jones (LJ) potential, often expressed in the form $V_{LJ}(r) = 4\epsilon [ (\sigma/r)^{12} - (\sigma/r)^{6} ]$, has been so far employed and parametrized in order to fit observations from specific experimental setups \cite{Kawai2016, Lodge2016}.
However, the oversimplified nature of the LJ potential, with just $\epsilon$ and $\sigma$ setting the two-body dissociation energy and equilibrium length, respectively, makes it extremely challenging to fit more than a few among the several physical quantities impacting the static and dynamical properties of a gold/graphene assembly.
As a consequence, previous studies have employed very different parametrizations to match some particular quantities of interest, with $\epsilon$ varying from $2.5$ to $90.0$ meV and $\sigma$ varying from $2.5$ to $3.2$ \AA\ \cite{Luedtke1999, Lewis2000, Sule2015, Kawai2016, Lodge2016}, making a straightforward comparison among these studies often impractical.

Beside quantitative errors, even a qualitative description in LJ terms is questionable. For a single gold atom residing on a graphene surface, typical pairwise isotropic potential (i.e., depending only on the distance between pairs of atoms), such as the LJ potential, would find the minimum energy position where the gold atom is located over a graphene hexagon center (hollow position). Nevertheless, experiments and first-principle calculations have demonstrated that the atop position, where the Au atom resides over a C atom, is energetically favorable \cite{Varns2008, Chen2016}. This exemplifies the need for an anisotropic potential to describe the gold/graphene interfacial energy landscape.

Beyond the single-atom contact, when an extended gold/graphene interface is considered, the sliding energy landscape becomes much more complex due to the intrinsic incommensurability of the contact.
In fact, nanoscale graphene flakes and ribbons residing over the (111) surface of gold can be found in both the epitaxially aligned R0 and the R30 tilted orientations \cite{Wofford2012, Kawai2016}.
Extended contacts, on the other hand, were predicted to prefer intermediate misfit angles $0<\theta<30^\circ$ \cite{Novaco1977}.
This, in turn, is expected to be manifested in the interfacial  tribological properties, where the combination of the weak Au-C interactions and the strong internal cohesive forces of gold and graphene may lead to exceptionally low friction coefficients ($<$$10^{-3}$) -- a condition often referred to as structural superlubricity \cite{Lodge2016, Yaniv2019, Li2020}.
In superlubric systems the corrugation energy (CE) -- the energy barrier resisting sliding -- can decrease to values much below the meV per interface atom. Therefore, studies aiming at describing superlubricity, often considering a combination of metals and graphitic materials \cite{Guerra2010, Camilli2014, Kawai2016, Gigli2017, Gigli2018, Trillitzsch2018, Auwarter2019}, are critically sensitive to the chosen experimental or theoretical setup.

It is therefore evident that there is an urgent need to develop a new force field able to reliably describe the interaction between graphene and gold. Unfortunately, there is only a handful of experimental studies providing data that can be directly employed to parametrize such a force field.
For example, Torres et al.\ measured an adhesion energy (AE) $E_a = 7687.1$\,mJ$\cdot$m$^{-2}$ in the case of graphene-covered gold nanoparticles \cite{Torres2017}, while a pull-off force of $F_a = 45.7 \pm 5.1$\,nN was measured by Li et al.\ for gold-coated AFM probes forming a $\sim200$\,nm$^2$ contact with a graphite substrate \cite{Li2020}. Other experiments provide only indirect information based on empirically fitted models or simulations \cite{Nie2012, Lodge2016, Kawai2016, Gigli2018}.
The comparison with experiments is further complicated by the presence of surface reconstruction at the gold surface, usually neglected in first-principles calculations due to the large supercell size required to encompass its long ($\sim6.3$\,nm) wavelength. Leaving out such reconstruction effects potentially overestimates the computed interaction energies, but in most cases this approximation yields only minor structural modifications in the model systems\cite{Hanke2013}.

Due to the lack of direct experimental measurements and accurate computational reference data for the adhesion and corrugation energies of graphene/gold interfaces, we performed dispersion corrected density functional theory (DFT+D3) calculations on the R30 tilted graphene/gold heterojunction. This reference dataset allowed us to parametrize a newly-developed semi-anisotropic interfacial potential (SAIP) that is able to simultaneously reproduce the DFT+D3 adhesion energy curves and sliding potential energy surfaces (PES). The developed SAIP and its suggested parameterization present a significant advancement with respect to the present stand of classical description of the interfacial interactions in gold and graphene junctions. Furthermore, the SAIP formulation provides a general tool for describing interfaces formed between two-dimensional (2D) materials and bulk solids.

\begin{figure}[bt!]
  \centering
  \includegraphics[width=\columnwidth]{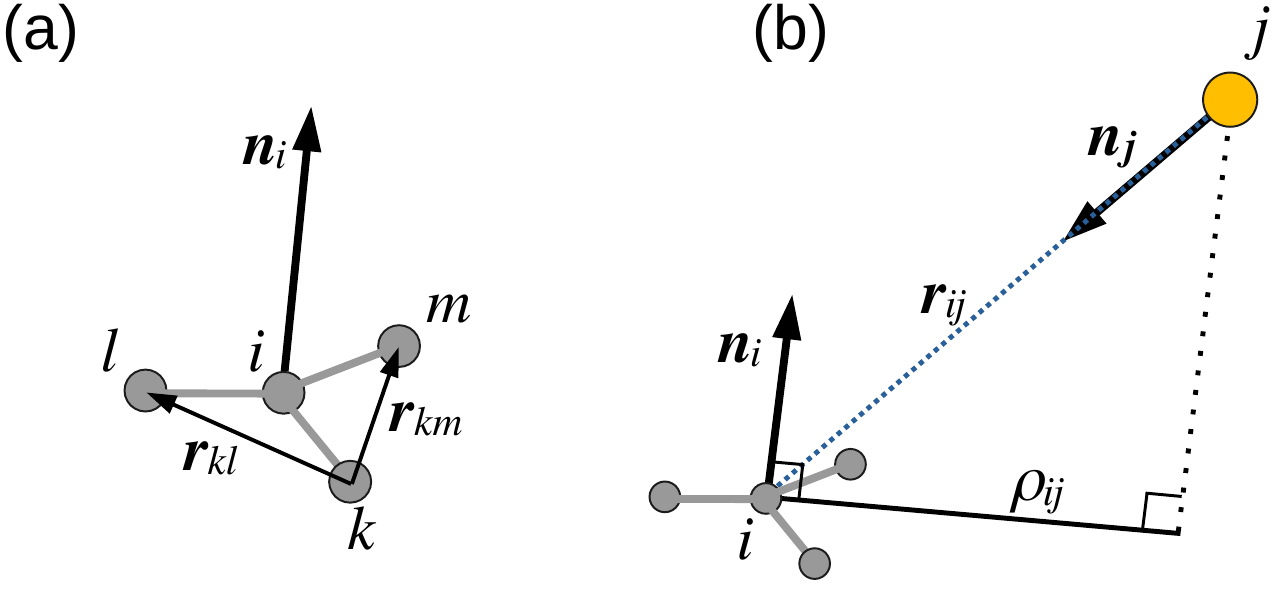}
  \caption{\small
  {\bf Normal vectors and transverse distance} -- (a) Construction of a normal vector $n_i$ associated with carbon or hydrogen atom $i$ (see text); (b) scheme of the relation between the normal vector $n_i$ and the transverse distance $\rho_{ij}$ between atom $i$ and gold atom $j$.}
  \label{fig.normals}
\end{figure}

\section{Potential Description}
⎄
The presented SAIP is based on the concept of anisotropic interlayer potentials (ILP) for 2D materials \cite{Kolmogorov2005,Leven2014,Leven2016,Maaravi2017,Ouyang2020}. The potential consists of two terms: an isotropic term that describes the long-range attractive dispersive interactions, and an anisotropic term that describes the Pauli-type repulsion between the graphene $\pi$ electrons and the gold surface electron density.
The dispersive term treats long-range van der Waals interactions via a $C_6/r^6$ LJ type potential, dampened in the short range with a Fermi-Dirac type function similar to that introduced in dispersion corrected DFT calculations to avoid double counting of correlation effects \cite{Tkatchenko2009}:

\begin{equation}\label{eq.E_dis}
 E_{dis}(r_{ij}) = \text{Tap}(r_{ij}) \left\{
  -\frac{1}
  { 1 + e^{-d \left [ \left( r_{ij}/\left(s_R \cdot r_{ij}^{eff}\right)\right) - 1\right] } } \cdot \frac{ C_{6,ij} }{ r_{ij}^6 } \right\}~~.
\end{equation}

Here, $r_{ij}$ is the distance between carbon or hydrogen atom $i$ and gold atom $j$, $d$ and $s_R$ are unit-less parameters determining the steepness and onset of the short-range Fermi-type dampening function, and $r_{ij}^{eff}$ and $C_{6,ij}$ are the sum of effective atomic radii and the pair-wise dispersion coefficients, respectively. The $\text{Tap}(r_{ij})$ function provides a continuous (up to 3$^{rd}$ derivative) long-range cutoff at $r_{ij} = R_{cut}$ to the potential aiming to reduce computational burden:\cite{deVosBurchart1992}

\begin{equation}\label{eq.Tap}
 \text{Tap}(r_{ij}) = \frac{20}{R_{cut}^7}r_{ij}^7 - \frac{70}{R_{cut}^6}r_{ij}^6 + \frac{84}{R_{cut}^5}r_{ij}^5 - \frac{35}{R_{cut}^4}r_{ij}^4 +1 ~~.
\end{equation}

Following the Kolmogorov-Crespi (KC) \cite{Kolmogorov2005} scheme, the anisotropic term of the potential is constructed from a Morse-like exponential isotropic term, multiplied by an anisotropic correction in which the orientation of graphene is described by normal vectors associated to each carbon or hydrogen atom:

\begin{equation}\label{eq.E_kc}
\begin{split}
E_{ani}(\bm{r}_{ij}) = \text{Tap}(r_{ij}) e^{ \alpha_{ij}\left( 1-\frac{r_{ij}}{\beta_{ij}} \right) }\times ~~~~~~~~~~~~~~~\\
  ~~~~~~~~\times \left[ \epsilon_{ij} + C \left( e^{-\left(\frac{\rho_{ij}}{\gamma_{ij}} \right)^2} + e^{-\left(\frac{\rho_{ji}}{\gamma_{ji}} \right)^2}  \right) \right]~.
\end{split}
\end{equation}

Here, $\text{Tap}(r_{ij})$ is the cutoff smoothing function of eq~\ref{eq.Tap}, $\alpha_{ij}$ and $\beta_{ij}$ set the slope and range of the potential, and $\gamma_{ij}$ sets the width of the Gaussian decay factors in the anisotropic correction term and thus determines the sensitivity to the transverse distance, $\rho_{ij}$, between carbon or hydrogen atom $i$ and gold atom $j$ (see Figure~\ref{fig.normals}). $C$ and $\epsilon_{ij}$ are constant scaling factors bearing units of energy.
The normalized normal vectors $\bm{n}_i$ (i.e., $\lVert \bm{n}_i \rVert \text{ = 1}$) serve to calculate the transverse distance $\rho_{ij}$ between pairs of carbon or hydrogen atom ($i$) and gold atom ($j$),
\begin{equation}\label{eq.rho}
\begin{split}
 \rho_{ij}^2 = r_{ij}^2 - (\bm{n}_i \cdot \bm{r}_{ij})^2 \\
 \rho_{ji}^2 = r_{ij}^2 - (\bm{n}_j \cdot \bm{r}_{ij})^2
\end{split}
\end{equation}
Each normal vector $\bm{n}_i$ defines the local normal direction to the graphene sheet (or to the benzene molecule) at the position of its atom $i$ (see Figure~\ref{fig.normals}). Note that the definition of the normal vector is not unique, and can follow different schemes \cite{Kolmogorov2005}. For example, one can calculate the normal vector of atom $i$ by averaging the three normalized cross products of the vectors connecting atom $i$ to its three nearest neighbors $k$,$l$,$m$, or in a more simple way, by calculating it as $\bm{n}_i = (\bm{r}_{kl}\times\bm{r}_{km})/(\lVert\bm{r}_{kl}\rVert {\,} \lVert \bm{r}_{km}\rVert)$ [see Figure~\ref{fig.normals}(a)].
In our implementation the former definition is adopted. However, when dealing with a flat graphene (or benzene molecule) lying on the $xy$ plane, all definitions of the normal vector should give $\bm{n}_i = \hat{z}$, making the specific choice irrelevant. In the case of curved graphene, such as for nanotubes, all normal vector definitions should produce very similar results except for extremely small radii of curvature \cite{Kolmogorov2005,Ouyang2018}.

To account for the isotropic nature of the isolated gold atoms electron cloud, their corresponding normal vectors are assumed to lie along the interatomic vector $\bm{r}_{ij}$.
Notably, this assumption is suitable for many bulk materials surfaces, e.g.\ for systems possessing s-type valence orbitals or metallic surfaces, whose valence electrons are mostly delocalized, such that their Pauli repulsions with the electrons of adjacent surfaces are isotropic. Caution should be used in the case of very small gold contacts, e.g.\ nanoclusters, where edge effects may become relevant.

Following the above assumption, $\bm{n}_{j}\|\bm{r}_{ij}$, one gets $\rho_{ji} \text{ = 0}$. The latter simplification reduces the anisotropic term to the following final form:

\begin{equation}\label{eq.E_ani}
\begin{split}
 E_{ani}(\bm{r}_{ij}) = \text{Tap}(r_{ij}) e^{ \alpha_{ij}\left( 1-\frac{r_{ij}}{\beta_{ij}} \right) }\times ~~~~~~~~~~~~~~~\\
 ~~~~~~~~\times \left[ \epsilon_{ij} + C \left( e^{-\left(\frac{\rho_{ij}}{\gamma_{ij}} \right)^2} + 1  \right) \right]~.
\end{split}
\end{equation}

\section{Model Systems and Methods}\label{sec.system_method}

\begin{figure}[bt!]
  \centering
  \includegraphics[width=\columnwidth]{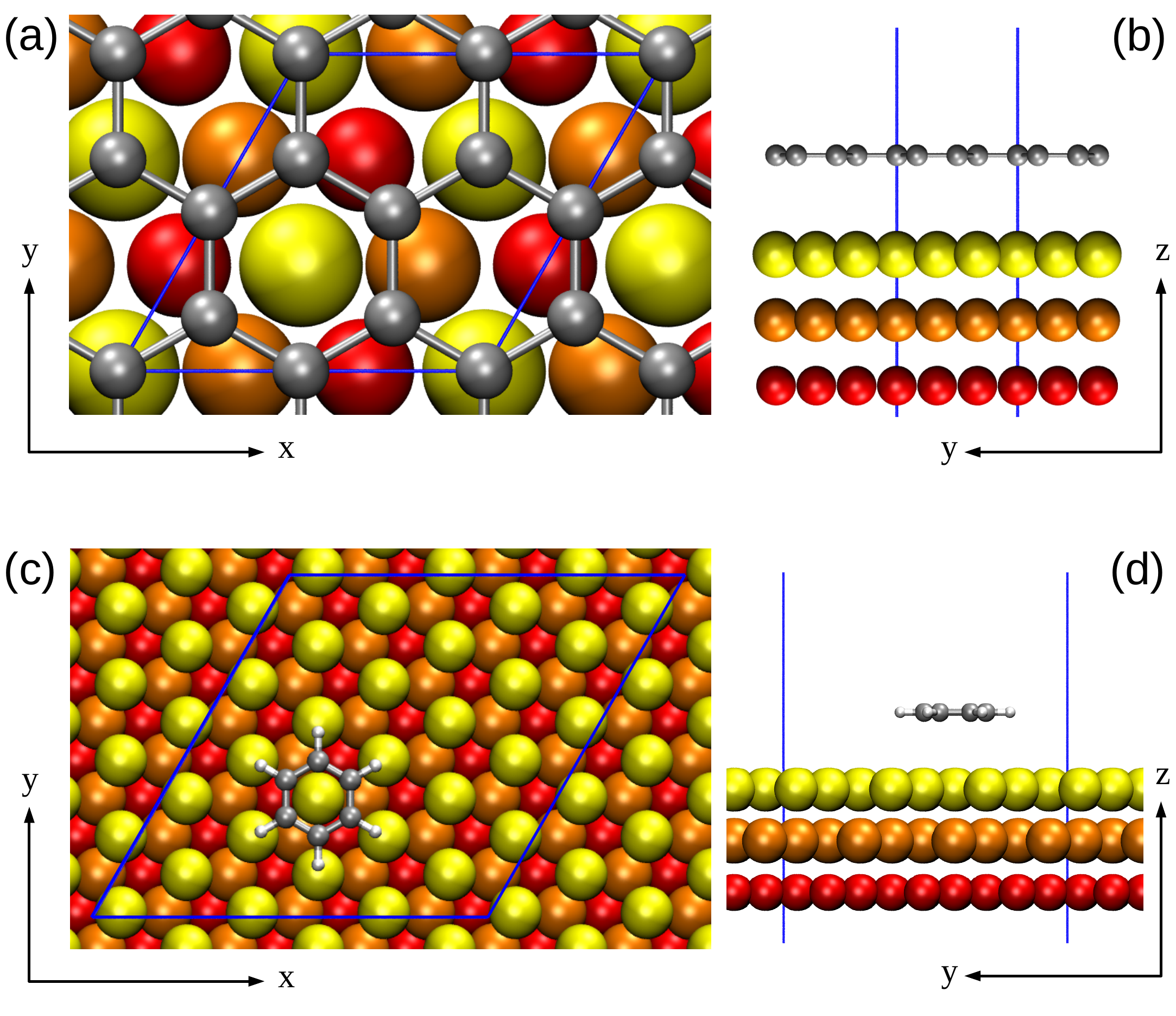}
  \caption{\small
  {\bf Reference model systems} -- (a) Top and (b) side views of the graphene/Au(111) model system in the 'atop' configuration. (c) Top and (d) side views of the benzene/Au(111) model system in the 'hollow' configuration. Carbon and hydrogen atoms are depicted in gray and white, respectively. First, second, and third Au layers are colored in yellow, orange, and red, respectively. Blue lines outline the primitive cell.}
  \label{fig.system}
\end{figure}

\subsection{Model systems}\label{sec.model}

As model systems for the parameterization process we choose the two interfaces depicted in Figure~\ref{fig.system}. The first is composed of a graphene layer positioned at the R30 stacking configuration over three Au(111) layers \cite{Wofford2012}. The second is composed of a benzene molecule residing over three Au(111) layers.
DFT spin-polarized calculations were performed by the Perdew-Burke-Ernzerhof (PBE) exchange-correlation functional within the generalized gradient approximation (GGA) augmented by Grimme's D3 long-range dispersion correction and the Rappe-Rabe-Kaxiras-Joannopoulos (RRKJ) ultra-soft core-corrected pseudopotentials, as implemented in the Quantum Espresso software.\cite{Perdew1996,Grimme2010,Rappe1990,Giannozzi2017}
Previous benchmark DFT calculations of metal–organic frameworks (MOFs) indicate that the PBE exchange correlation density functional approximation with the D3 dispersion correction can accurately describe the energetics of complex systems involving organic and inorganic components \cite{Grimme2010,Nazarian2015,Mostaani2015}. However, whether this is the best DFT approach (out of tens of currently available dispersion oriented density functional approximations) for the system under consideration cannot currently be concluded, mainly due to the lack of experimental results or high-accuracy computational reference data.

Kinetic-energy cutoffs of $600$\,eV for the wavefunctions and $4952$\,eV for the density were employed, with a reciprocal-space mesh of $7$\,$\times$\,$7$\,$\times$\,$1$ k-points for the Au(111)/graphene interface and $3$\,$\times$\,$3$\,$\times$\,$1$ k-points for the Au(111)/benzene system (see below). Convergence of total energy over the above cutoffs and k-points grid was established.
Periodic boundary conditions were applied in all directions. In all the calculations graphene or benzene were kept flat \cite{Tesch2016} and the gold atoms were maintained in the fcc bulk configuration.

In the Au(111)/graphene case, the lateral stress was minimized by iteratively rescaling the system size. The final supercell vectors length and angles were $a=b=4.965638$\,\AA, $c=30$\,\AA, $\alpha=\beta=90^\circ$, $\gamma=60^\circ$ (see Figure~\ref{fig.system}a-b).
The out-of-plane periodicity $c$ was chosen to be as large as possible to avoid interactions with replicas along the vertical direction. The calculated lattice constants of the isolated bulk gold and graphene systems were $2.8825$ and $2.4652$~\AA, corresponding to a final strain of $-0.54$\% and $+0.7$\%, respectively, in the relaxed composite system.
In the Au(111)/benzene and Au(111)/C$_{60}$ cases we have employed a supercell with $a=b=14.896914$\,\AA, $c=30$\,\AA, $\alpha=\beta=90^\circ$, $\gamma=60^\circ$ (see Figure~\ref{fig.system}c-d).
Equilibrium distances of $3.5$\,\AA\ between graphene and gold, of $3.3$\,\AA\ between benzene and gold, and of $3.3$\,\AA\ between C$_{60}$ and gold were found by rigid vertical displacement of the adsorbate molecule.

\begin{table*}[tb!]
  \caption{\small
  {\bf Potential parameters} --
    List of SAIP (eqs \ref{eq.E_dis} and \ref{eq.E_ani}) parameters for the interfacial Au-C and Au-H interactions.
  }
\begin{center}
{
\renewcommand{\arraystretch}{1.5}  
\setlength{\tabcolsep}{2pt}        
\begin{tabular}{cccccccccc}
  \hline
  \hline
   atoms~~ & $\alpha$ &$\beta$\,(\AA) & $\gamma$\,(\AA) & $\epsilon$\,(meV) & $C$\,(meV) & $d$ & $s_R$ & $r_{\text{eff}}$\,(\AA) & $C_6$\,($\text{eV} \cdot \text{\AA}^6$)\\
  \hline
   Au--C~~ & 13.56556 & 3.69133 &  1.01755  & 7.09648   & -1.03683   & 11.05865 & 1.06356  & 3.75526 & 81.58471 \\
   Au--H~~ & 4.30650  & 3.78996 &  10.68118 & 225.08878 & -111.68915 & 18.61492 & 0.983319 & 3.35076 & 70.68654 \\
  \hline
  \hline
\end{tabular}
}
\end{center}
\label{tab.parameters}
\end{table*}

\begin{figure}[b!]
  \centering
  \newcommand\x{1.0}
  \hspace{\fpeval{0.15*\x}\columnwidth}
  \includegraphics[width=\fpeval{0.56*\x}\columnwidth]{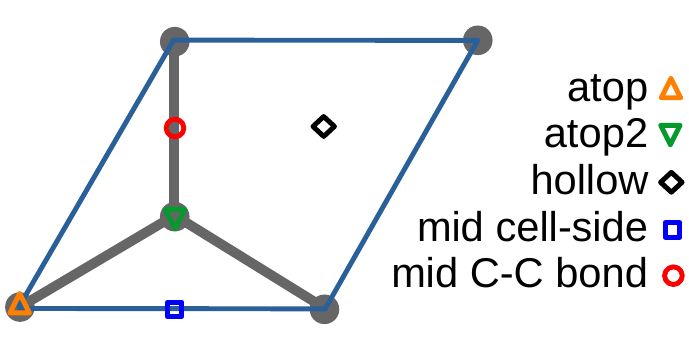}\\
  \includegraphics[width=\fpeval{0.75*\x}\columnwidth]{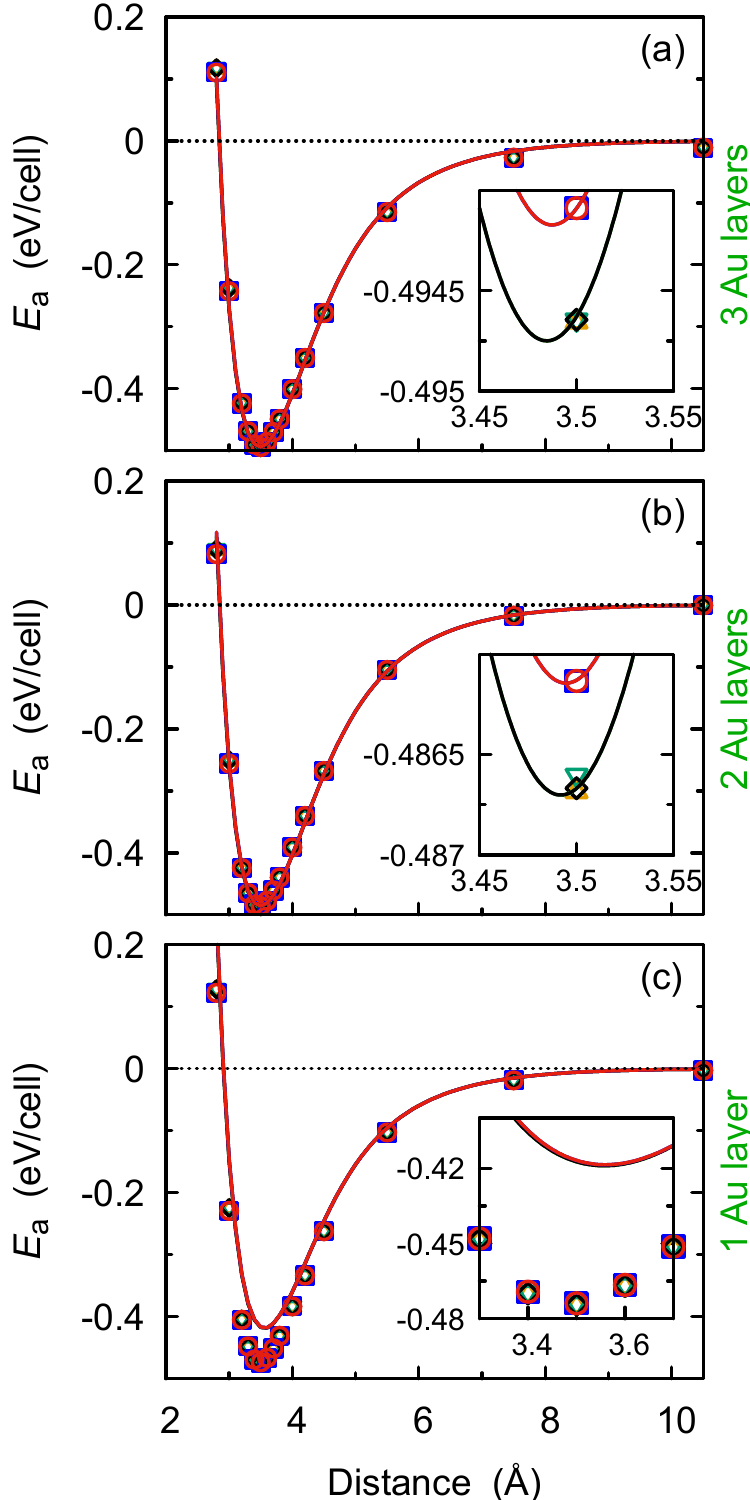}
  \caption{\small
  {\bf Graphene on gold: adhesion} --
  Adhesion energy curves of the graphene/gold heterostructure with (a) trilayer gold, (b) bilayer gold, and (c) single layer gold. Symbols and lines represent DFT+D3 reference data and SAIP results, respectively. Insets provide a magnification around the minimum energy. Different colors of symbols and lines represent different stacking modes, as depicted in the top panel (see text).}
  \label{fig.adhesion}
\end{figure}

\begin{figure*}[tb!]
  \centering
  \includegraphics[width=\textwidth]{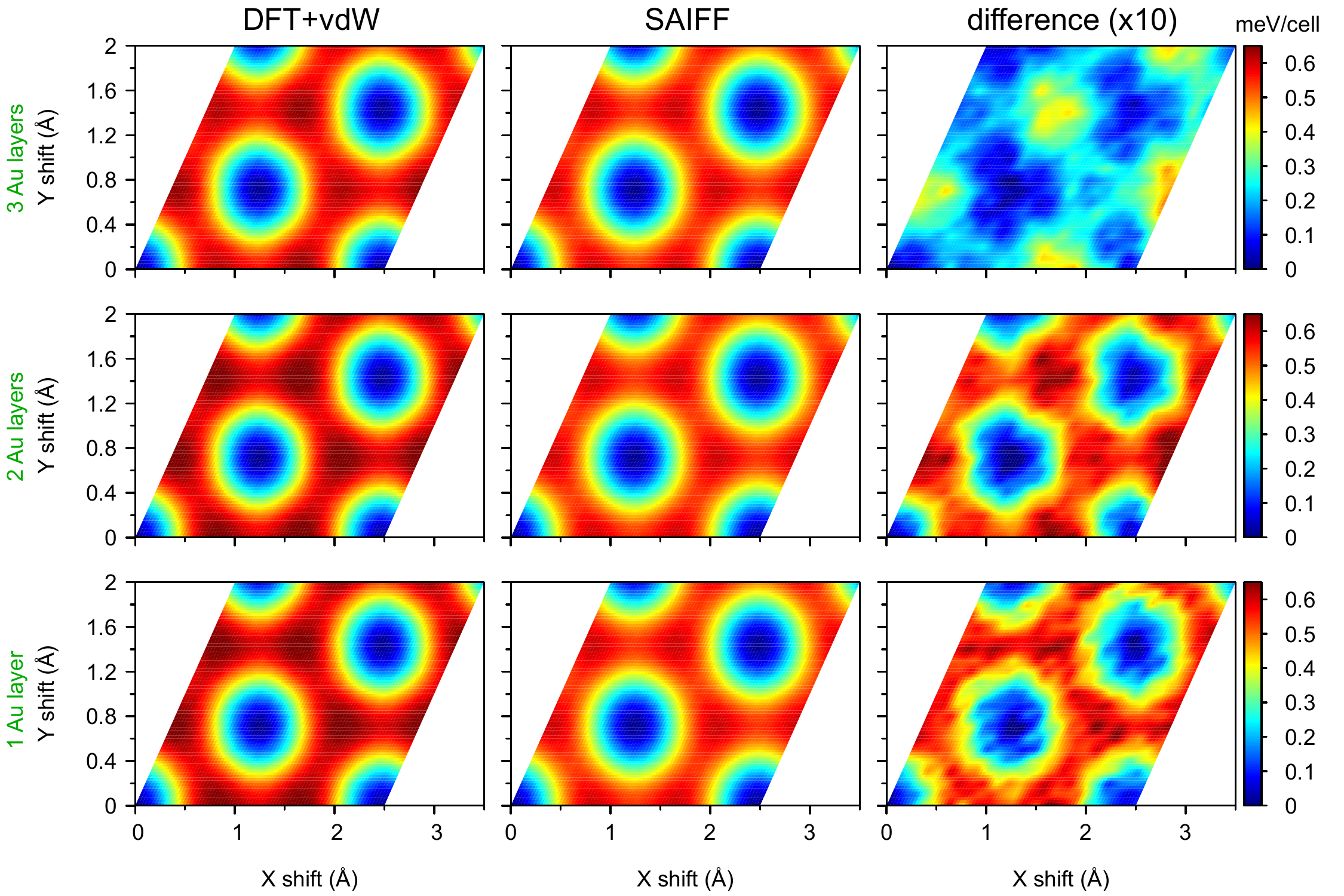}
  \caption{\small
  {\bf Graphene on gold: PES} --
  Sliding potential energy surfaces (PES) of the graphene/gold heterostructure, calculated at an interface separation of 3.5\,\AA. The upper, middle, and bottom rows are for trilayer, bilayer, and single layer gold, respectively. The left, center, and right columns present the PES calculated using DFT+D3, SAIP, and their difference, respectively. In the latter, the differences are magnified $\times10$ to clearly present the fine features. For better visibility a 4$\times$4 linear interpolation has been applied to all the PES maps. The minimum of the PES was shifted to zero for clarity.}
  \label{fig.pes}
\end{figure*}

\begin{figure*}[tb!]
  \centering
  \includegraphics[width=0.7\textwidth]{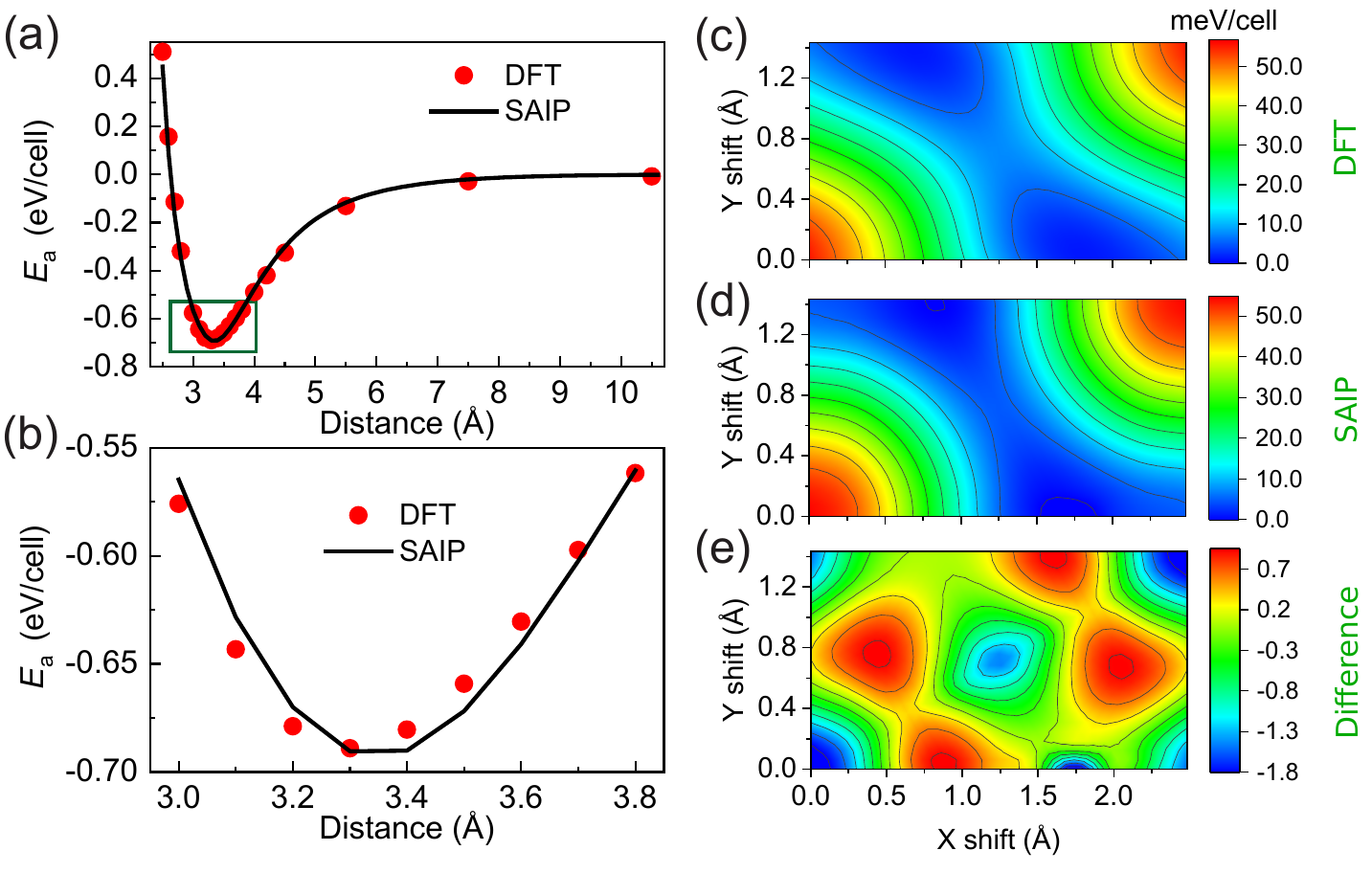}
  \caption{\small
  {\bf Benzene on gold} --
  (a) Adhesion energy curve, and (b) a zoom-in around its minimum, of a benzene molecule residing over a trilayer gold surface at the configuration depicted in Figure~\ref{fig.system}c and d. Symbols and lines represent DFT+D3 reference data and SAIP results, respectively. The corresponding sliding PESs calculated using (c) DFT+D3, (d) SAIP, and (e) their difference are obtained at a benzene-gold separation of 3.3\,\AA. For better visibility, isoenergetic contour lines are superposed on the PES maps, which were smoothened by a 8$\times$8 linear interpolation. The minimum of the PES was shifted to zero for clarity.}
  \label{fig.benzene}
\end{figure*}

\subsection{Fitting protocol}\label{sec.fitting}

The parameters of the SAIP were determined against reference $M$ = $M_b$ + $M_s$ DFT datasets including $M_b$ AE curves and $M_s$ sliding PESs. The AE curves were calculated for five high-symmetry stacking modes (see top panel of Figure~\ref{fig.adhesion}), which are concisely denoted by $\bm{r}_{m}$, where $m \in [1,M_b]$, such that $\bm{r}_{m} \in \mathbb{R}^{3N_{m}}$ and $N_{m}$ is the number of atoms in configuration $m$. Each AE curve includes $15$ data points as a function of the gold-graphene (gold-benzene, or gold-fullerene) distance. The sliding PESs were obtained at a fixed vertical ($z$) distance of $3.5$\,\AA, by rigidly shifting the adsorbate along the lateral ($x$-$y$) direction with respect to the gold substrate. The single-point total energy at each of $441$ points of an uniform mesh grid was recorded. The origin (0,0) configurations of the PESs correspond to those presented in Figure~\ref{fig.system}.

Optimal SAIP parameters were obtained by minimizing the following objective function that quantifies the difference between the DFT reference data and the potential predictions:

\begin{equation}\label{eq.obj}
\begin{split}
 \it{\Phi}(\bm{\xi}) = \sum_{m\text{=1}}^{M_b} w_{m}^{b} \left\lVert \bm{E}_{m}^{\text{b}}(\bm{r}_{m},\bm{\xi}) - \bm{E}_{m}^{\text{b,DFT}} \right\rVert _\mathrm{2} \\
 + \sum_{m\text{=1}}^{M_s} \ w_{m}^{s} \left\lVert \bm{E}_{m}^{\text{s}}(\bm{r}_{m},\bm{\xi}) - \bm{E}_{m}^{\text{s,DFT}} \right\rVert _\mathrm{2}~~.
\end{split}
\end{equation}

Here, $\lVert \cdot \rVert _\text{2}$ is the  Euclidean norm (2-norm) that measures the difference between the SAIP predictions and the DFT reference data, $\bm{\xi}$ represents the set of potential parameters, and $\bm{E}_{m}^{\text{b}}(\bm{r}_{m},\bm{\xi})$ and $\bm{E}_{m}^{\text{s}}(\bm{r}_{m},\bm{\xi})$ represent the $M_b$ AE curves and $M_s$ sliding PES data sets, respectively. $w_{m}^{b}$ and $w_{m}^{s}$ are the corresponding weighting coefficients.
The reference DFT interfacial energies, $\bm{E}_{m}^{\text{b,DFT}}$ and $\bm{E}_{M}^{\text{s,DFT}}$, are obtained as follows: For any given configuration $m$ of the heterostructure, the total energy is first obtained from DFT+D3 calculation: $\bm{E}_{m}^{\text{DFT,total}}$. Then the energy of the isolated graphene, $\bm{E}_{m}^{\text{DFT,graphene}}$, and of isolated gold, $\bm{E}_{m}^{\text{DFT,gold}}$, are computed separately using the same cell as that of the composite system. The DFT interfacial energy appearing in eq~\ref{eq.obj} is then defined as:

\begin{equation}\label{eq.Ebind1}
 \bm{E}_{m}^{\text{b/s,DFT}} = \bm{E}_{m}^{\text{DFT,total}} - {E}_{m}^{\text{DFT,graphene}} - {E}_{m}^{\text{DFT,gold}}~~.
\end{equation}

Since the AE -- defined as the difference between maximum and minimum PES energy -- is much smaller than the AE ($\sim$1 vs.\ hundreds of meV/cell) the energy weights for the AE curves were set to $w_{m}^{b}=1$ ($m$\,=\,1,\dots,5), and those of the sliding PESs were chosen as $w_{M_s}^{s}=100$, thus providing comparable precision. The optimization was carried out using MATLAB with an interior-point algorithm \cite{Byrd2000, Waltz2006} (further details are provided in Refs.\ \cite{Ouyang2018, Ouyang2020}). To obtain transferable parameters that can account for varying gold thickness, we first parameterized the potential for the heterostructure with trilayer gold using the objective function defined in eq~\ref{eq.obj}, then we added the training sets of the heterostructure with bilayer and single layer gold and reparameterized the potential following the same procedure.

\section{Results and Discussion}\label{sec.discussion}

The fitted parameters for Au-C and Au-H interactions are reported in Table ~\ref{tab.parameters}. We note that the negative sign in the $C$ parameter can be attributed to the fact that the atop position of an Au atom on graphene is energetically favorable with respect to the hollow position \cite{Varns2008,Chen2016}, where the Au atom resides over the graphene hexagon center (see top panel of  Figure~\ref{fig.adhesion}). When applying this potential to describe the interfacial interaction in other 2D and bulk material interfaces, the parameter $C$ can be either positive or negative, depending on the sliding energetics \cite{Kolmogorov2005,Ouyang2018}.

\subsection{Graphene on Gold}\label{sec.graphene}

Figure~\ref{fig.adhesion} shows the comparison between the AE curves of the Au(111)/graphene heterostructure with trilayer (Figure~\ref{fig.adhesion}a), bilayer (Figure~\ref{fig.adhesion}b) and single layer gold (Figure~\ref{fig.adhesion}c) obtained using DFT+D3 (symbols) and the SAIP using the parameters provided in Table~\ref{tab.parameters} (solid lines), calculated at different stacking modes.
Good agreement between the DFT and SAIP is obtained especially in the tri- and bi-layer case, while for the single layer a larger discrepancy is found.
Note that the DFT calculations show a similar AE regardless of the number of gold layers, while the SAIP predicts a reduced adhesion in the case of a single gold layer.
We associate the former with the fact that decreasing the number of layers reduces the adhesive interactions, but at the same time the under coordinated Au surface becomes more chemically reactive. This is also related to the known increased atomic density of gold surfaces with respect to bulk \cite{Yang2015,Wang2019,Wang2020,Zhu2020}. Such compensation yields an almost unchanged AE value, just slightly reduced with respect to the 2-layers case. This change of reactivity is lacking in the present form of the SAIP. Here, the reduced adhesion obtained in the case of the gold monolayer is clearly due to the reduced number of Au-C interacting pairs, whereas the binding energies obtained for the 2- and 3-layer substrates are very similar being the third layer distant from the graphene surface.
The average deviation, $\left[\sum_{\text{1}}^{N} \frac{1}{N} \left( \bm{E}^{\text{SAIP}} - \bm{E}^{\text{DFT}} \right)^2\right]^{1/2}$, for the AE curves of the tri-, bi-, and single-layer Au(111) systems are $11.38$, $11.43$, and $70.95$ meV/cell, respectively.

As mentioned earlier, comparison with experiment is challenging due to the limited availability of measured AE values for the graphene/gold interface. Specifically, Torres et al.\ found an AE of $E_a = 7687.1$\,mJ$\cdot$m$^{-2} = 48$\,eV/nm$^2$ in an experiment exploring graphene-covered gold nanoparticles \cite{Torres2017}, whereas Li et al.\ found a pull-off force of ${\approx}0.23$\,nN/nm$^2 = 1.44$\,eV/nm$^3$ for gold-graphite interfaces \cite{Li2020}. Notably, in order to obtain the AE measure by Torres et al.\, this force (even if assumed to remain constant) has to be applied along a distance of $\sim 33$\,nm, way beyond the interlayer interaction range. This indicates a discrepancy between the two experimentally measured values. We note that our calculated AE of about $0.5$\,eV/cell ($375$\,mJ$\cdot$m$^{-2}$) obtained for the trilayer gold/graphene interface model (Fig.~\ref{fig.adhesion}a) is lower than that found by Torres et al.\ while the corresponding pull-off force of $2.04$\,nN/nm$^2$, evaluated from the first derivative of the SAIP AE curves, is larger than the experimental value of Li et al., suggesting that the SAIP provides values within the experimentally available range.
A much better agreement is obtained between the SAIP predictions and previous computational results, such as the AE of $467$\,mJ$\cdot$m$^{-2}$ calculated by Tesch et al.\ for graphene nanoflakes on Au(111) \cite{Tesch2016}, and $394$\,mJ$\cdot$m$^{-2}$ recently calculated for the Cu(111)/graphene interface \cite{Han2019} that is also dominated by vdW interactions.

The sliding PESs calculated using the SAIP parametrization show good agreement with the reference DFT+D3 data (see Figure~\ref{fig.pes}), as well, with an error of 7.8\%, 11\%, and 10\% for the PES corrugation of the trilayer, bilayer, and single layer gold, respectively,
and corresponding average deviations of $0.010$, $0.020$, and $0.044$ meV/cell, respectively.
When comparing the AE of the different stacking modes considered (see Fig.~\ref{fig.adhesion}a) one finds that the maximum difference is about $0.5$\,meV/cell. The latter value is related to the CE as shown in Fig.~\ref{fig.pes}.
We note that the CE is of the order of $0.075$\,meV per C-atom, indicating an exceptionally lubric contact \cite{Yaniv2019}.
We would like to note that we obtain CE and AE values using energy differences of nearly identical systems, thus we expect a very high numerical accuracy.
Furthermore, in the case of rigid contacts considered herein, such small CE ensues from the approximation employed to describe the lattice mismatch between gold and graphene. A truly incommensurate lattice mismatch, only obtainable in the thermodynamic limit of an infinite supercell size, would in fact yield a vanishing CE regardless of the potential parameters.

A LJ fitting of the DFT+D3 tri-layer gold/ graphene PES resulted in $\epsilon=14.86$\,meV and $\sigma=3.8$\,\AA, in agreement with previous semi-empirical LJ parameterizations used to reproduce experimental results of gold clusters diffusion on graphene \cite{Luedtke1999, Lodge2016}. Notably, a LJ fitting of the AE curves results in substantially different values, $\epsilon=8.50$\,meV and $\sigma = 3.37$\,\AA, confirming that the isotropic LJ description is incapable of simultaneously describing the binding and sliding physics of vdW interfaces.

\begin{figure*}[tb!]
  \centering
  \includegraphics[width=0.45\textwidth]{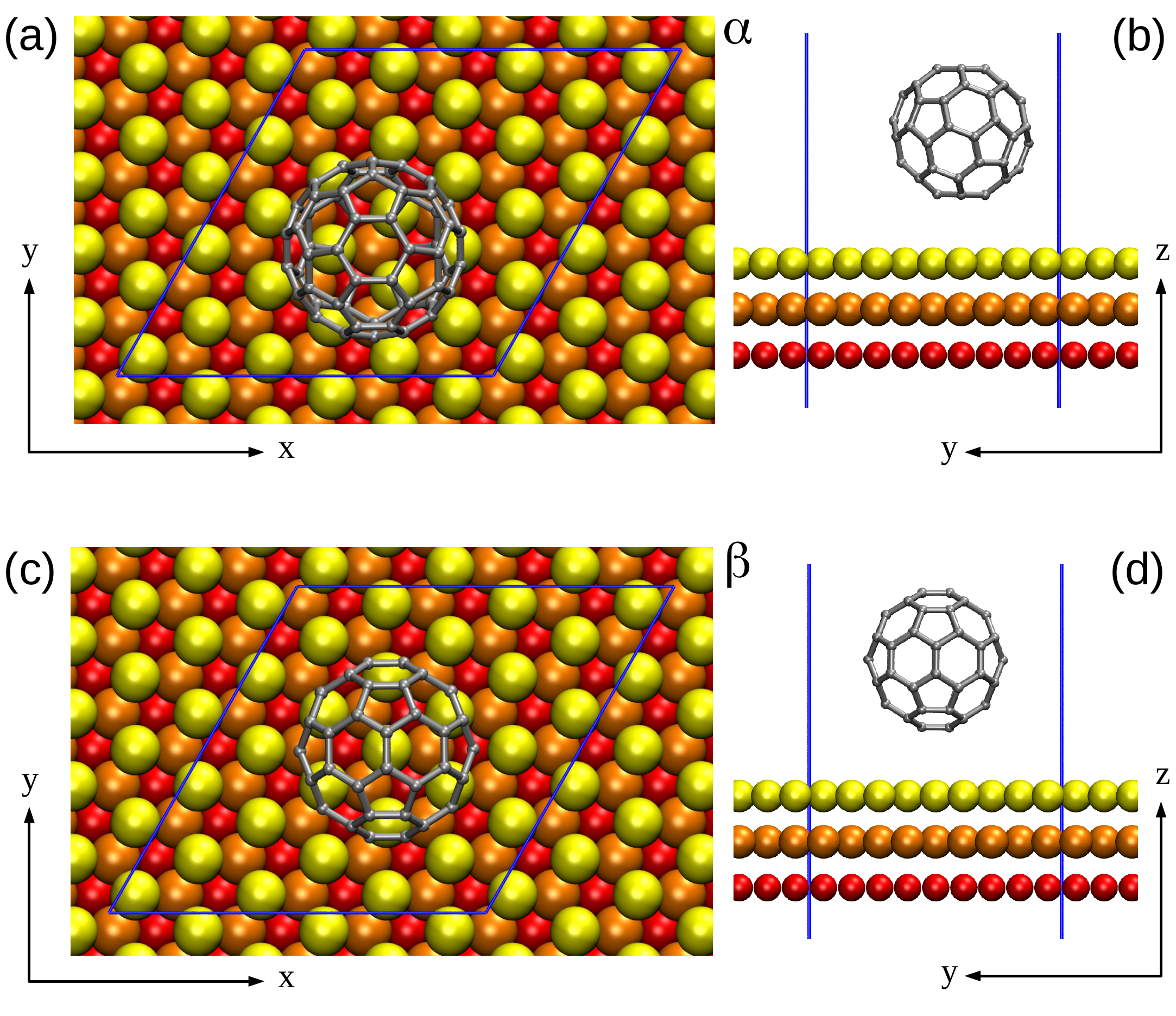}~~~~~~~~
  \includegraphics[width=0.45\textwidth]{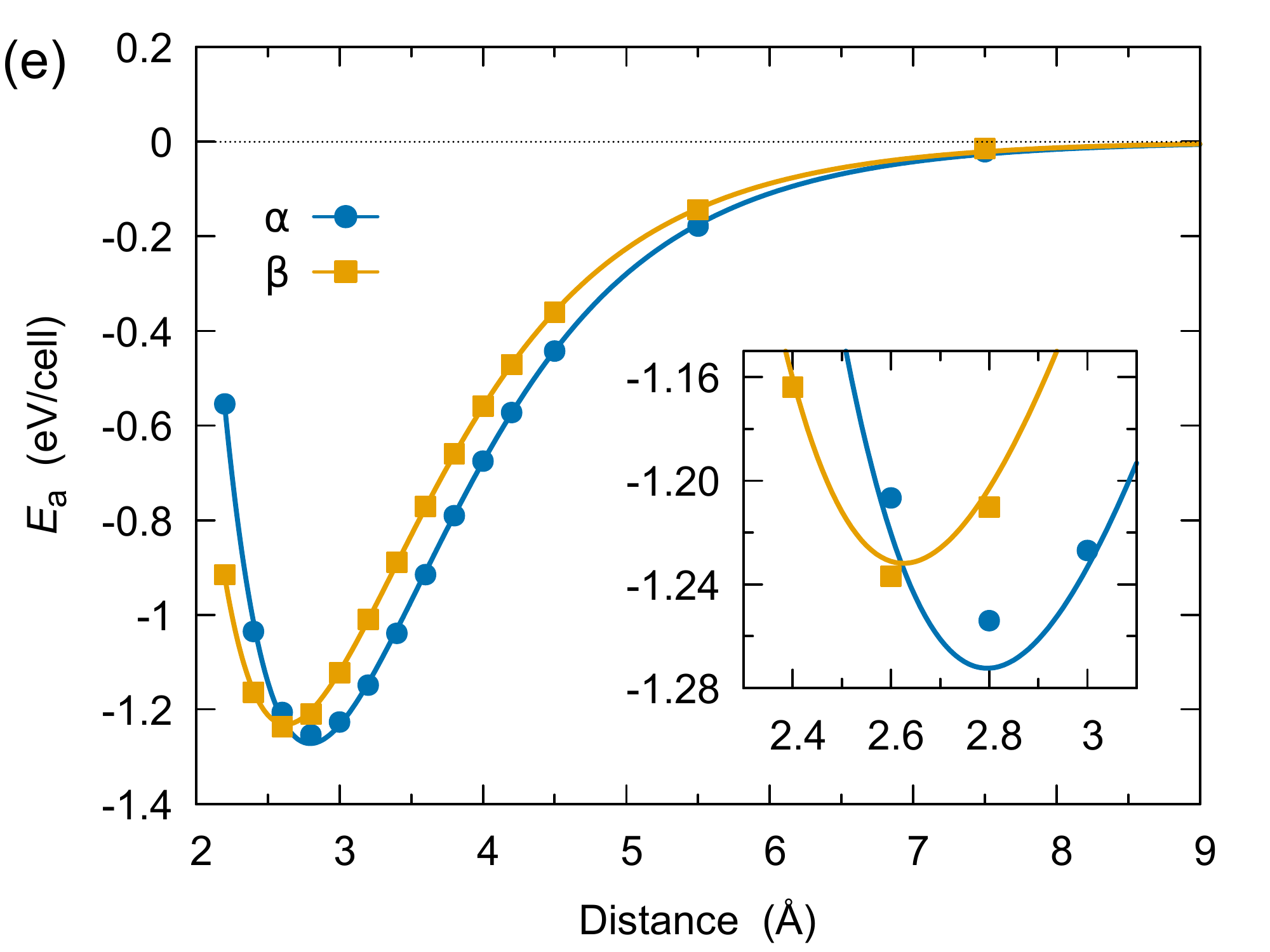}
  \caption{\small
  {\bf C$_{60}$ on gold} -- (a) Top and (b) side views of the C$_{60}$/Au(111) model system in the $\alpha$ configuration. (c) Top and (d) side views of the C$_{60}$/Au(111) model system in the $\beta$ configuration. Carbon atoms are depicted in gray, and first, second, and third Au layers are colored in yellow, orange, and red, respectively. Blue lines outline the primitive cell.
  (e) Adhesion energy curves for the $\alpha$ and $\beta$ configurations, with symbols corresponding to the reference DFT+D3 data, and solid lines to SAIP results obtained using the parameters appearing in Table~\ref{tab.c60}.}
  \label{fig.c60}
\end{figure*}

\begin{table*}[tb!]
  \caption{\small
  {\bf C$_{60}$ parameters} --
    List of SAIP (eqs \ref{eq.E_dis} and \ref{eq.E_ani}) parameters for the C$_{60}$--gold interaction.
  }
\begin{center}
{
\renewcommand{\arraystretch}{1.5}  
\setlength{\tabcolsep}{2pt}        
\begin{tabular}{cccccccccc}
  \hline
  \hline
   atoms~~ & $\alpha$ & $\beta$\,(\AA) & $\gamma$\,(\AA) & $\epsilon$\,(meV) & $C$\,(meV) & $d$ & $s_R$ & $r_{\text{eff}}$\,(\AA) & $C_6$\,($\text{eV} \cdot \text{\AA}^6$)\\
  \hline
   Au--C~~ & 11.25629 & 3.42336 & 5.37200 & 0.186756 & -0.093183 & 8.89407 & 1.098183 & 3.650286 & 90.433089 \\
  \hline
  \hline
\end{tabular}
}
\end{center}
\label{tab.c60}
\end{table*}

\subsection{Benzene on gold}\label{sec.benzene}

Going beyond the periodic interface, we next consider the case of a benzene molecule residing atop a gold surface. Figure~\ref{fig.benzene} reports the AE curves (Figure~\ref{fig.benzene}a,b) and sliding PES (Figure~\ref{fig.benzene}c-e) of the benzene/Au(111) heterostructure obtained using DFT+D3 and the SAIP using the parameters provided in Table~\ref{tab.parameters}. Good agreement between the DFT and SAIP results is found for both the AE curve and the sliding PES,
with average deviations of $28.0$\,meV/cell ($\sim4$\% of the AE) and $0.73$\,meV/cell ($\sim1.4$\% of the CE) for the AE curve and the PES, respectively, and a deviation of 1.7\% in the overall PES corrugation. The obtained AE value of $0.689$\,eV is in good agreement with a reference experimental value of $\sim0.64$\,eV at finite temperature.\cite{Syomin2001}

The SAIP parameterization yields a smaller Au-H $C_6$ coefficient and a larger $d$ value than those of the corresponding Au-C parameters, suggesting a weaker and shorter-range dispersion term for the former. Furthermore, the Au-H anisotropic repulsive term (eq~\ref{eq.E_ani}) is considerably weaker than the corresponding Au-C term. This results from the tenfold larger $\gamma$ parameter of the former (see Table~\ref{tab.parameters}) that yields $e^{-\left(\frac{\rho_{ij}}{\gamma_{ij}}\right)}\sim1$ for any reasonable value of $\rho_{ij}$ in eq~\ref{eq.E_ani}. Together with the fact that $\epsilon \simeq 2|C|$ in the Au-H case (see Table~\ref{tab.parameters}) the square brackets appearing in eq~\ref{eq.E_ani} are small in magnitude and quite insensitive to the value of the lateral interatomic distance. This indicates that the Au-H interaction has minor effect on the lateral shear motion and that most of the lateral forces originate from the Au-C interaction. Therefore, we expect that the present parametrization should hold for other planar benzenoid systems.

\subsection{C$_{60}$ on gold}\label{sec.c60}

Finally, we challenge our developed SAIP against extremely bent graphitic systems. In this case, one might expect significant deviations of the interfacial energy profiles from the ideally flat graphene case discussed above. In particular, it is known that mechanical bending can alter the reactivity of graphene sheets \cite{Park2003}, and introduce additional interfacial effects, such as curvature-induced structural lubricity \cite{Zande2019}. Since the latter effects should mainly depend on the structural properties (geometry) of the adjacent surfaces, they should be captured by the SAIP with the parameterization presented in Table~\ref{tab.parameters}. However, accounting for the different reaction energy of the curved graphene requires a full quantum-mechanical treatment, and thus a reparameterization of the SAIP for the curved system.
To demonstrate this, we considered the case of a C$_{60}$ fullerene physisorbed over an Au(111) surface in two different orientations, marked as $\alpha$ and $\beta$, as depicted in Figure~\ref{fig.c60} (a)-(d).
By performing DFT+D3 calculations at different separations from the Au(111) surface we obtained the adhesion curves, reported in Figure~\ref{fig.c60}e, showing adhesion energies of 1.2545\,eV and 1.2373\,eV for the $\alpha$ and $\beta$ configurations, respectively.
On average, the C$_{60}$ AE is larger than that of benzene by about 80\%, thus supporting the need to provide a separate parameterization for curved systems.
This is in agreement with previous calculations indicating an increased reaction energy for NTs of smaller curvature radius \cite{Park2003}.
The increased surface reactivity is manifested also in the equilibrium distance from the gold surface, which reduces from $3.48$\,\AA\ for flat graphene to $2.6$ and $2.8$ \AA\ for C$_{60}$ at the $\alpha$ and $\beta$ configurations, respectively (see Figure~\ref{fig.c60}e).
The simultaneous fitting of the two adhesion curves for C$_{60}$ produced a distinct set of SAIP parameters, reported in Table~\ref{tab.c60}.
In comparison with the graphene/Au(111) interface, the larger $C_6$ and smaller $d$ parameters indicate a stronger and longer range vdW dispersion term. Nonetheless, the anisotropic term reduces significantly, mostly due to a $\epsilon \simeq 2|C|$ balance (see discussion above), indicating that the Au-C interaction is weekly dependent on shear motion in this highly curved system and that the sliding energy surface corrugation is expected to be very small.
Similar considerations could be made in the case of very small-diameter carbon nanotubes (CNTs) on gold, for which Table~\ref{tab.c60} could apply. However, the SAIP parameters should approach those of Table~\ref{tab.parameters} as the CNT diameter increases. The possibility of including such parameter transition in a revised SAIP functional form capable of describing general CNT/gold assemblies\cite{Singh2010} is presently under investigation.

\section{Conclusions}\label{sec.conclusions}

The results presented above indicate that the proposed semi-anisotropic interface potential (SAIP) is able to accurately reproduce the energetics of graphene-gold, benzene-gold, and C$_{60}$-gold interactions, as obtained from DFT+D3 calculations.
While the functional form of the SAIP is suitable to treat many interfaces between graphitic systems and gold surfaces, system specific parametrizations of the SAIP are recommended in order to obtain optimal accuracy.
However, the presented potential is expected to describe the structural and dynamical response to external forces of a large number of prototypical systems, such as gold nanoclusters on graphite \cite{Guerra2010, Dietzel2013}, graphene nanoribbons on Au(111) \cite{Kawai2016, Gigli2017, Gigli2018}, and C$_{60}$ on Au(111) \cite{Schull2007, Shin2014}, among others, with much improved accuracy with respect to previous classical models.
Furthermore, our formulation can be generalized to describe a wide variety of interfaces between hexagonal two-dimensional materials and bulk solids, such as MoS$_2$/Au \cite{Trillitzsch2018}, {\it h}-BN/Au \cite{Camilli2014, Auwarter2019}, or graphene/Ag \cite{Tesch2016}. This, in turn, will considerably increase the scope of material interfaces that can be treated using reliable dedicated classical force-fields.

\section*{Acknowledgments}
We acknowledge fruitful discussions with (in alphabetical order): Neil Drummond, Leeor Kronik, Ryo Maezono, Kida Noriko, Alexandre Tkatchenko, and Keishu Utimula.
W.O.\ acknowledges financial support from the starting-up fund of Wuhan University and the National Natural Science Foundation of China (nos.\ 11890673 and 11890674).
O.H.\ is grateful for the generous financial support of the Israel Science Foundation under grant no.\,1586/17, Tel Aviv University Center for Nanoscience and Nanotechnology, and the Naomi Foundation for generous financial support via the 2017 Kadar Award.
R.G.\ acknowledges financial support from Universit\`a degli Studi di Milano, grant no.\,1094 SEED 2020 - TEQUAD, support by the Italian Ministry of University and Research through PRIN UTFROM no.\ 20178PZCB5, and computational support from the Center for Complexity and Biosystems.

\section*{Authors Contributions}
RG designed the research and performed the DFT calculations. WO fitted the parameters, coded the potential into LAMMPS, and performed the MD simulations. WO, OH, and RG participated in the force-field development, scientific discussions, and writing of the paper.

\section*{Competing Interests}
The authors have no competing interest to declare.

\bibliographystyle{unsrt}
\bibliography{biblio}

\begin{thebibliography}{10}

\bibitem{Luedtke1999}
W.~D. Luedtke and Uzi Landman.
\newblock Slip diffusion and l{\'{e}}vy flights of an adsorbed gold
  nanocluster.
\newblock {\em Physical Review Letters}, 82(19):3835--3838, May 1999.

\bibitem{Lewis2000}
Laurent~J. Lewis, Pablo Jensen, Nicolas Combe, and Jean-Louis Barrat.
\newblock Diffusion of gold nanoclusters on graphite.
\newblock {\em Physical Review B}, 61(23):16084--16090, June 2000.

\bibitem{Guerra2010}
Roberto Guerra, Ugo Tartaglino, Andrea Vanossi, and Erio Tosatti.
\newblock Ballistic nanofriction.
\newblock {\em Nature Materials}, 9(8):634--637, July 2010.

\bibitem{Dietzel2013}
Dirk Dietzel, Michael Feldmann, Udo~D. Schwarz, Harald Fuchs, and Andr{\'{e}}
  Schirmeisen.
\newblock Scaling laws of structural lubricity.
\newblock {\em Physical Review Letters}, 111(23), December 2013.

\bibitem{Cihan2016}
Ebru Cihan, Semran {\.{I}}pek, Engin Durgun, and Mehmet~Z. Baykara.
\newblock Structural lubricity under ambient conditions.
\newblock {\em Nature Communications}, 7(1), June 2016.

\bibitem{Lodge2016}
M.~S. Lodge, C.~Tang, B.~T. Blue, W.~A. Hubbard, A.~Martini, B.~D. Dawson, and
  M.~Ishigami.
\newblock Lubricity of gold nanocrystals on graphene measured using quartz
  crystal microbalance.
\newblock {\em Scientific Reports}, 6(1), August 2016.

\bibitem{Kawai2016}
S.~Kawai, A.~Benassi, E.~Gnecco, H.~Sode, R.~Pawlak, X.~Feng, K.~Mullen,
  D.~Passerone, C.~A. Pignedoli, P.~Ruffieux, R.~Fasel, and E.~Meyer.
\newblock Superlubricity of graphene nanoribbons on gold surfaces.
\newblock {\em Science}, 351(6276):957--961, February 2016.

\bibitem{Gigli2017}
L~Gigli, N~Manini, A~Benassi, E~Tosatti, A~Vanossi, and R~Guerra.
\newblock Graphene nanoribbons on gold: understanding superlubricity and edge
  effects.
\newblock {\em 2D Materials}, 4(4):045003, August 2017.

\bibitem{Gigli2018}
Lorenzo Gigli, Shigeki Kawai, Roberto Guerra, Nicola Manini, R{\'{e}}my Pawlak,
  Xinliang Feng, Klaus M\"{u}llen, Pascal Ruffieux, Roman Fasel, Erio Tosatti,
  Ernst Meyer, and Andrea Vanossi.
\newblock Detachment dynamics of graphene nanoribbons on gold.
\newblock {\em {ACS} Nano}, 13(1):689--697, December 2018.

\bibitem{Muszynski2008}
Ryan Muszynski, Brian Seger, and Prashant~V. Kamat.
\newblock Decorating graphene sheets with gold nanoparticles.
\newblock {\em The Journal of Physical Chemistry C}, 112(14):5263--5266, March
  2008.

\bibitem{Zhu2013}
Xiaolong Zhu, Lei Shi, Michael~S. Schmidt, Anja Boisen, Ole Hansen, Jian Zi,
  Sanshui Xiao, and N.~Asger Mortensen.
\newblock Enhanced light{\textendash}matter interactions in graphene-covered
  gold nanovoid arrays.
\newblock {\em Nano Letters}, 13(10):4690--4696, September 2013.

\bibitem{Turcheniuk2015}
Kostiantyn Turcheniuk, Rabah Boukherroub, and Sabine Szunerits.
\newblock Gold{\textendash}graphene nanocomposites for sensing and biomedical
  applications.
\newblock {\em Journal of Materials Chemistry B}, 3(21):4301--4324, 2015.

\bibitem{Goncalves2009}
Gil Goncalves, Paula A. A.~P. Marques, Carlos~M. Granadeiro, Helena I.~S.
  Nogueira, M.~K. Singh, and J.~Gra\'cio.
\newblock Surface modification of graphene nanosheets with gold nanoparticles:
  The role of oxygen moieties at graphene surface on gold nucleation and
  growth.
\newblock {\em Chemistry of Materials}, 21(20):4796--4802, October 2009.

\bibitem{Song2019}
Yongchao Song, Tailin Xu, Li-Ping Xu, and Xueji Zhang.
\newblock Nanodendritic gold/graphene-based biosensor for tri-mode {miRNA}
  sensing.
\newblock {\em Chemical Communications}, 55(12):1742--1745, 2019.

\bibitem{Tesch2016}
Julia Tesch, Philipp Leicht, Felix Blumenschein, Luca Gragnaniello, Mikhail
  Fonin, Lukas Eugen~Marsoner Steinkasserer, Beate Paulus, Elena Voloshina, and
  Yuriy Dedkov.
\newblock Structural and electronic properties of graphene nanoflakes on
  au(111) and ag(111).
\newblock {\em Scientific Reports}, 6(1), March 2016.

\bibitem{Forti2020}
Stiven Forti, Stefan Link, Alexander St\"{o}hr, Yuran Niu, Alexei~A. Zakharov,
  Camilla Coletti, and Ulrich Starke.
\newblock Semiconductor to metal transition in two-dimensional gold and its van
  der waals heterostack with graphene.
\newblock {\em Nature Communications}, 11(1), May 2020.

\bibitem{Sule2015}
Péter S\"{u}le, Márton Szendrő, Gábor~Zsolt Magda, Chanyong Hwang, and
  Levente Tapasztó.
\newblock Nanomesh-type graphene superlattice on au(111) substrate.
\newblock {\em Nano Letters}, 15(12), 2015.

\bibitem{Varns2008}
R~Varns and P~Strange.
\newblock Stability of gold atoms and dimers adsorbed on graphene.
\newblock {\em Journal of Physics: Condensed Matter}, 20(22):225005, April
  2008.

\bibitem{Chen2016}
Qu~Chen, Kuang He, Alex~W. Robertson, Angus~I. Kirkland, and Jamie~H. Warner.
\newblock Atomic structure and dynamics of epitaxial 2d crystalline gold on
  graphene at elevated temperatures.
\newblock {\em {ACS} Nano}, 10(11):10418--10427, November 2016.

\bibitem{Wofford2012}
Joseph~M Wofford, Elena Starodub, Andrew~L Walter, Shu Nie, Aaron Bostwick,
  Norman~C Bartelt, Konrad Th\"{u}rmer, Eli Rotenberg, Kevin~F McCarty, and
  Oscar~D Dubon.
\newblock Extraordinary epitaxial alignment of graphene islands on au(111).
\newblock {\em New Journal of Physics}, 14(5):053008, May 2012.

\bibitem{Novaco1977}
Anthony~D. Novaco and John~P. McTague.
\newblock Orientational epitaxy{\textemdash}the orientational ordering of
  incommensurate structures.
\newblock {\em Physical Review Letters}, 38(22):1286--1289, May 1977.

\bibitem{Yaniv2019}
Rotem Yaniv and Elad Koren.
\newblock Robust superlubricity of gold{\textendash}graphite heterointerfaces.
\newblock {\em Advanced Functional Materials}, 30(18):1901138, April 2019.

\bibitem{Li2020}
Jinjin Li, Jianfeng Li, Xinchun Chen, Yuhong Liu, and Jianbin Luo.
\newblock Microscale superlubricity at multiple gold{\textendash}graphite
  heterointerfaces under ambient conditions.
\newblock {\em Carbon}, 161:827--833, May 2020.

\bibitem{Camilli2014}
L~Camilli, E~Sutter, and P~Sutter.
\newblock Growth of two-dimensional materials on non-catalytic substrates:
  h-{BN}/au(111).
\newblock {\em 2D Materials}, 1(2):025003, August 2014.

\bibitem{Trillitzsch2018}
Felix Trillitzsch, Roberto Guerra, Arkadiusz Janas, Nicola Manini, Franciszek
  Krok, and Enrico Gnecco.
\newblock Directional and angular locking in the driven motion of au islands on
  {MoS}2.
\newblock {\em Physical Review B}, 98(16), October 2018.

\bibitem{Auwarter2019}
Willi Auw\"{a}rter.
\newblock Hexagonal boron nitride monolayers on metal supports: Versatile
  templates for atoms, molecules and nanostructures.
\newblock {\em Surface Science Reports}, 74(1):1--95, 2019.

\bibitem{Torres2017}
Jorge Torres, Yisi Zhu, Pei Liu, Seong~Chu Lim, and Minhee Yun.
\newblock Adhesion energies of 2d graphene and {MoS}2 to silicon and metal
  substrates.
\newblock {\em physica status solidi (a)}, 215(1):1700512, December 2017.

\bibitem{Nie2012}
Shu Nie, Norman~C. Bartelt, Joseph~M. Wofford, Oscar~D. Dubon, Kevin~F.
  McCarty, and Konrad Th\"{u}rmer.
\newblock Scanning tunneling microscopy study of graphene on au(111): Growth
  mechanisms and substrate interactions.
\newblock {\em Physical Review B}, 85(20), May 2012.

\bibitem{Hanke2013}
Felix Hanke and Jonas Bj\"{o}rk.
\newblock Structure and local reactivity of the au(111) surface reconstruction.
\newblock {\em Physical Review B}, 87(23):235422, June 2013.

\bibitem{Kolmogorov2005}
Aleksey~N. Kolmogorov and Vincent~H. Crespi.
\newblock Registry-dependent interlayer potential for graphitic systems.
\newblock {\em Physical Review B}, 71(23), June 2005.

\bibitem{Leven2014}
Itai Leven, Ido Azuri, Leeor Kronik, and Oded Hod.
\newblock Inter-layer potential for hexagonal boron nitride.
\newblock {\em The Journal of Chemical Physics}, 140(10), May 2014.

\bibitem{Leven2016}
Itai Leven, Tal Maaravi, Ido Azuri, Leeor Kronik, and Oded Hod.
\newblock Interlayer potential for graphene/h-bn heterostructures.
\newblock {\em Journal of Chemical Theory and Computation}, 12(6), May 2016.

\bibitem{Maaravi2017}
Tal Maaravi, Itai Leven, Ido Azuri, Leeor Kronik, and Oded Hod.
\newblock Interlayer potential for homogeneous graphene and hexagonal boron
  nitride systems: Reparametrization for many-body dispersion effects.
\newblock {\em The Journal of Physical Chemistry C}, 121(41), 2017.

\bibitem{Ouyang2020}
Wengen Ouyang, Ido Azuri, Davide Mandelli, Alexandre Tkatchenko, Leeor Kronik,
  Michael Urbakh, and Oded Hod.
\newblock Mechanical and tribological properties of layered materials under
  high pressure: Assessing the importance of many-body dispersion effects.
\newblock {\em Journal of Chemical Theory and Computation}, 16(1), 2020.

\bibitem{Tkatchenko2009}
Alexandre Tkatchenko and Matthias Scheffler.
\newblock Accurate molecular van der waals interactions from ground-state
  electron density and free-atom reference data.
\newblock {\em Physical Review Letters}, 102(7), February 2009.

\bibitem{deVosBurchart1992}
E.~de~Vos~Burchart, V.A. Verheij, H.~van Bekkum, and B.~van~de Graaf.
\newblock A consistent molecular mechanics force field for all-silica zeolites.
\newblock {\em Zeolites}, 12(2):183--189, February 1992.

\bibitem{Ouyang2018}
Wengen Ouyang, Davide Mandelli, Michael Urbakh, and Oded Hod.
\newblock Nanoserpents: Graphene nanoribbon motion on two-dimensional hexagonal
  materials.
\newblock {\em Nano Letters}, 18(9), 2018.

\bibitem{Perdew1996}
John~P. Perdew, Kieron Burke, and Matthias Ernzerhof.
\newblock Generalized gradient approximation made simple.
\newblock {\em Physical Review Letters}, 77(18):3865--3868, October 1996.

\bibitem{Grimme2010}
Stefan Grimme, Jens Antony, Stephan Ehrlich, and Helge Krieg.
\newblock A consistent and accurate ab initio parametrization of density
  functional dispersion correction ({DFT}-d) for the 94 elements h-pu.
\newblock {\em The Journal of Chemical Physics}, 132(15):154104, April 2010.

\bibitem{Rappe1990}
Andrew~M. Rappe, Karin~M. Rabe, Efthimios Kaxiras, and J.~D. Joannopoulos.
\newblock Optimized pseudopotentials.
\newblock {\em Physical Review B}, 41(2):1227--1230, January 1990.

\bibitem{Giannozzi2017}
P~Giannozzi, O~Andreussi, T~Brumme, O~Bunau, M~Buongiorno Nardelli, M~Calandra,
  R~Car, C~Cavazzoni, D~Ceresoli, M~Cococcioni, N~Colonna, I~Carnimeo, A~Dal
  Corso, S~de~Gironcoli, P~Delugas, R~A DiStasio, A~Ferretti, A~Floris,
  G~Fratesi, G~Fugallo, R~Gebauer, U~Gerstmann, F~Giustino, T~Gorni, J~Jia,
  M~Kawamura, H-Y Ko, A~Kokalj, E~K\"{u}{\c{c}}\"{u}kbenli, M~Lazzeri,
  M~Marsili, N~Marzari, F~Mauri, N~L Nguyen, H-V Nguyen, A~Otero de-la Roza,
  L~Paulatto, S~Ponc{\'{e}}, D~Rocca, R~Sabatini, B~Santra, M~Schlipf, A~P
  Seitsonen, A~Smogunov, I~Timrov, T~Thonhauser, P~Umari, N~Vast, X~Wu, and
  S~Baroni.
\newblock Advanced capabilities for materials modelling with quantum
  {ESPRESSO}.
\newblock {\em Journal of Physics: Condensed Matter}, 29(46):465901, October
  2017.

\bibitem{Nazarian2015}
Dalar Nazarian, P.~Ganesh, and David~S. Sholl.
\newblock Benchmarking density functional theory predictions of framework
  structures and properties in a chemically diverse test set of
  metal{\textendash}organic frameworks.
\newblock {\em Journal of Materials Chemistry A}, 3(44):22432--22440, 2015.

\bibitem{Mostaani2015}
E.~Mostaani, N.{\hspace{0.167em}}D. Drummond, and V.{\hspace{0.167em}}I.
  Fal'ko.
\newblock Quantum monte~carlo calculation of the binding energy of bilayer
  graphene.
\newblock {\em Physical Review Letters}, 115(11), September 2015.

\bibitem{Byrd2000}
Richard~H. Byrd, Jean~Charles Gilbert, and Jorger Nocedal.
\newblock A trust region method based on interior point techniques for
  nonlinear programming.
\newblock {\em Mathematical Programming}, 89(1), 2000.

\bibitem{Waltz2006}
R.A. Waltz, J.L. Morales, J.~Nocedal, and D.~Orban.
\newblock An interior algorithm for nonlinear optimization that combines line
  search and trust region steps.
\newblock {\em Mathematical Programming}, 107(3), 2006.

\bibitem{Yang2015}
Li-Ming Yang, Matthew Dornfeld, Thomas Frauenheim, and Eric Ganz.
\newblock Glitter in a 2d monolayer.
\newblock {\em Physical Chemistry Chemical Physics}, 17(39):26036--26042, 2015.

\bibitem{Wang2019}
Xuelu Wang, Chunyang Wang, Chunjin Chen, Huichao Duan, and Kui Du.
\newblock Free-standing monatomic thick two-dimensional gold.
\newblock {\em Nano Letters}, 19(7):4560--4566, June 2019.

\bibitem{Wang2020}
T.~Wang, M.~Park, Q.~Yu, J.~Zhang, and Y.~Yang.
\newblock Stability and synthesis of 2d metals and alloys: a review.
\newblock {\em Materials Today Advances}, 8:100092, December 2020.

\bibitem{Zhu2020}
Qi~Zhu, Youran Hong, Guang Cao, Yin Zhang, Xiaohan Zhang, Kui Du, Ze~Zhang,
  Ting Zhu, and Jiangwei Wang.
\newblock Free-standing two-dimensional gold membranes produced by extreme
  mechanical thinning.
\newblock {\em {ACS} Nano}, 14(12):17091--17099, November 2020.

\bibitem{Han2019}
Yong Han, King~C. Lai, Ann Lii-Rosales, Michael~C. Tringides, James~W. Evans,
  and Patricia~A. Thiel.
\newblock Surface energies, adhesion energies, and exfoliation energies
  relevant to copper-graphene and copper-graphite systems.
\newblock {\em Surface Science}, 685:48--58, July 2019.

\bibitem{Syomin2001}
Denis Syomin, Jooho Kim, Bruce~E. Koel, and G.~Barney Ellison.
\newblock Identification of adsorbed phenyl (c6h5) groups on metal
  surfaces:{\hspace{0.167em}} electron-induced dissociation of benzene on
  au(111).
\newblock {\em The Journal of Physical Chemistry B}, 105(35):8387--8394, August
  2001.

\bibitem{Park2003}
Seongjun Park, Deepak Srivastava, and Kyeongjae Cho.
\newblock Generalized chemical reactivity of curved surfaces:~ carbon
  nanotubes.
\newblock {\em Nano Letters}, 3(9):1273--1277, September 2003.

\bibitem{Zande2019}
Edmund Han, Jaehyung Yu, Emil Annevelink, Jangyup Son, Dongyun~A. Kang, Kenji
  Watanabe, Takashi Taniguchi, Elif Ertekin, Pinshane~Y. Huang, and Arend~M.
  van~der Zande.
\newblock Ultrasoft slip-mediated bending in few-layer graphene.
\newblock {\em Nature Materials}, 19(3):305--309, November 2019.

\bibitem{Singh2010}
Rajpal Singh, Thathan Premkumar, Ji-Young Shin, and Kurt{\hspace{0.25em}}E.
  Geckeler.
\newblock Carbon nanotube and gold-based materials: A symbiosis.
\newblock {\em Chemistry - A European Journal}, 16(6):1728--1743, February
  2010.

\bibitem{Schull2007}
G.~Schull and R.~Berndt.
\newblock Orientationally ordered (7 x 7) superstructure of c60 on au(111).
\newblock {\em Physical Review Letters}, 99(22), November 2007.

\bibitem{Shin2014}
Heekeun Shin, A.~Schwarze, R.~D. Diehl, K.~Pussi, A.~Colombier,
  {\'{E}}.~Gaudry, J.~Ledieu, G.~M. McGuirk, L.~N.~Serkovic Loli,
  V.~Fourn{\'{e}}e, L.~L. Wang, G.~Schull, and R.~Berndt.
\newblock Structure and dynamics of c60 molecules on au(111).
\newblock {\em Physical Review B}, 89(24), June 2014.

\end{thebibliography}
\end{document}